\documentclass[pdflatex,sn-mathphys-num,Numbered]{sn-jnl}% Math and Physical Sciences Numbered Reference Style

\usepackage{graphicx}%
\usepackage{multirow}%
\usepackage{amsmath,amssymb,amsfonts}%
\usepackage{amsthm}%
\usepackage{mathrsfs}%
\usepackage[title]{appendix}%
\usepackage{xcolor}%
\usepackage{textcomp}%
\usepackage{manyfoot}%
\usepackage{booktabs}%
\usepackage{algorithm}%
\usepackage{algorithmicx}%
\usepackage{algpseudocode}%
\usepackage{listings}%
\usepackage{ulem}

\newcommand{\bs}{\boldsymbol}

\theoremstyle{thmstyleone}%

\theoremstyle{thmstyletwo}%

\theoremstyle{thmstylethree}%

\begin{document}

\title[Article Title]{A high-significance detection of primordial tidal torque imprints}

\author[1]{\fnm{Ming-Jie} \sur{Sheng}}

\author*[1]{\fnm{Hao-Ran} \sur{Yu}}\email{haoran@xmu.edu.cn}

\author[2,3]{\fnm{Min} \sur{Bao}}

\author[1]{\fnm{Bing-Hang} \sur{Chen}}

\author[1]{\fnm{Fang-Na} \sur{Shao}}

\author[4,5,6]{\fnm{Qi} \sur{Guo}}

\author[2,3]{\fnm{Yan-Mei} \sur{Chen}}

\author[7,8]{\fnm{Huiyuan} \sur{Wang}}

\author[9,10,11,12,13]{\fnm{Ue-Li} \sur{Pen}}

\author[5,4,6]{\fnm{Jie} \sur{Wang}}

\author[14,15]{\fnm{Xiaohu} \sur{Yang}}

\affil*[1]{\orgdiv{Department of Astronomy}, \orgname{Xiamen University}, \orgaddress{\city{Xiamen}, \postcode{361005}, \state{Fujian}, \country{China}}}

\affil[2]{\orgdiv{School of Astronomy and Space Science}, \orgname{Nanjing University}, \orgaddress{\city{Nanjing}, \postcode{210023}, \state{Jiangsu}, \country{China}}}

\affil[3]{\orgdiv{Key Laboratory of Modern Astronomy and Astrophysics (Nanjing University)}, \orgname{Ministry of Education}, \orgaddress{\city{Nanjing}, \postcode{210023}, \state{Jiangsu}, \country{China}}}

\affil[4]{\orgdiv{Institute for Frontiers in Astronomy and Astrophysics}, \orgname{Beijing Normal University}, \orgaddress{\city{Beijing}, \postcode{102206}, \country{China}}}

\affil[5]{\orgdiv{National Astronomical Observatories}, \orgname{Chinese Academy of Sciences}, \orgaddress{\city{Beijing}, \postcode{100101}, \country{China}}}

\affil[6]{\orgdiv{School of Astronomy and Space Science}, \orgname{University of Chinese Academy of Sciences (UCAS)}, \orgaddress{\city{Beijing}, \postcode{100049}, \country{China}}}

\affil[7]{\orgdiv{Department of Astronomy}, \orgname{University of Science and Technology of China}, \orgaddress{\city{Hefei}, \postcode{230026}, \state{Anhui}, \country{China}}}

\affil[8]{\orgdiv{School of Astronomy and Space Science}, \orgname{University of Science and Technology of China}, \orgaddress{\city{Hefei}, \postcode{230026}, \state{Anhui}, \country{China}}}

\affil[9]{\orgname{Academia Sinica Institute of Astronomy and Astrophysics (ASIAA)}, \orgaddress{\city{Taipei}, \postcode{10617}, \state{Taiwan}}}

\affil[10]{\orgdiv{Canadian Institute for Theoretical Astrophysics}, \orgname{University of Toronto}, \orgaddress{\city{Toronto}, \postcode{M5S 3H8}, \state{Ontario}, \country{Canada}}}

\affil[11]{\orgdiv{Dunlap Institute for Astronomy \& Astrophysics}, \orgname{University of Toronto}, \orgaddress{\city{Toronto}, \postcode{M5S 3H4}, \state{Ontario}, \country{Canada}}}

\affil[12]{\orgname{Perimeter Institute for Theoretical Physics}, \orgaddress{\city{Waterloo}, \postcode{N2L 2Y5}, \state{Ontario}, \country{Canada}}}

\affil[13]{\orgdiv{Canadian Institute for Advanced Research}, \orgname{CIFAR Program in Gravitation and Cosmology}, \orgaddress{\city{Toronto}, \postcode{M5G 1M1}, \state{Ontario}, \country{Canada}}}

\affil[14]{\orgdiv{State Key Laboratory of Dark Matter Physics, Tsung-Dao Lee Institute}, \orgname{Shanghai Jiao Tong University}, \orgaddress{\city{Shanghai}, \postcode{200240}, \country{China}}}

\affil[15]{\orgdiv{School of Physics and Astronomy}, \orgname{Shanghai Jiao Tong University}, \orgaddress{\city{Shanghai}, \postcode{200240}, \country{China}}}

\abstract{Tidal-torque theory predicts that galaxy angular momenta are imprinted by the primordial tidal field acting on proto-structures and that they can retain information about the early Universe through cosmic evolution. Here we test this prediction by comparing observed galaxy angular momentum vectors with those predicted from the primordial density field reconstructed by ELUCID for the nearby Universe. Among the galaxy populations considered, the gas component of central massive elliptical galaxies provides the clearest signal, exhibiting a strong direction correlation at a significance of about $7\sigma$. These results provide a robust observational evidence for tidal-torque theory and open a window for cosmological measurements of neutrino mass and other cosmological parameters.}

\maketitle

\section{Introduction}\label{sec1}
Angular momentum is a fundamental quantity governing galaxy formation and evolution. 
It regulates galaxy morphology, size, and internal kinematics and plays a key 
role in shaping the diversity of the observed galaxy population 	
\cite{1980MNRAS.193..189F,1998MNRAS.295..319M,2014ApJ...784...26O,2024JCAP...05..111M}. 
From a cosmological perspective, understanding how galaxies acquire and evolve their angular 
momenta is essential for linking present-day observables to the early Universe, one of the 
core tasks of large-scale structure studies \cite{2005MNRAS.360L..82R,2021JCAP...06..024M}. 
As a parity-odd vector quantity, galaxy angular momentum provides information inaccessible 
to scalar density fields by offering further degrees of freedom for probing cosmology and 
testing fundamental physics, such as neutrino masses \cite{2020ApJ...898L..27L}, potential 
parity violation \cite{2020PhRvL.124j1302Y,2025PhRvL.135n1002S}, dynamical dark energy 
\cite{2020ApJ...902...22L}, and large-scale axial intrinsic alignment \cite{2025arXiv251110005D}. 
It, thus, serves as a key bridge between the large-scale tidal environment that shapes dark 
matter (DM) halos and the small-scale baryonic processes that govern galaxy evolution.

In this context, the tidal-torque theory (TTT) \cite{1969ApJ...155..393P,1970Afz.....6..581D,
1984ApJ...286...38W} provides a physically motivated framework that explains the origin of 
galaxy angular momentum. 
TTT predicts that angular momentum is generated when large-scale tidal fields exert gravitational 
torques on protohalos --- the progenitors of present-day DM halos --- owing to the misalignment 
between their inertia tensors and the surrounding tidal field. 
These torques generate coherent angular momenta that persist until halos virialize 
\cite{2002MNRAS.332..325P}. 
At low redshift, the angular momenta of DM halos is still correlated with the primordial 
tidal field, both in direction \cite{2000ApJ...532L...5L,2001ApJ...555..106L,
2002MNRAS.332..325P} and magnitude \cite{2021PhRvD.103f3522W}. 
Galaxies forming within these halos inherit, at least partially, their angular momenta 
\cite{1980MNRAS.193..189F,1998MNRAS.295..319M}. 
Hydrodynamical simulations further show that the angular momenta of central galaxies tend 
to align with those of their host halos \cite{2010MNRAS.404.1137B}, and that both remain correlated 
with the primordial tidal environment \cite{2023ApJ...943..128S,2024PhRvD.109l3548S}, making 
galaxy angular momentum a sensitive probe of cosmic initial conditions.
Testing TTT observationally therefore provides a unique way for connecting galaxy-scale 
kinematics to the physics of structure formation. 

Although the rotation of DM halos cannot be observed directly, the measurable galaxy spins 
(hereafter we refer to the `angular momentum direction' as `spin' for brevity) 
provide an observational proxy for the primordial tidal field. 
At low redshift, the three-dimensional (3D) spins of spiral galaxies can be inferred using 
their ellipticities, projection angles, spiral-arm parities, Doppler asymmetries, 
and dust absorptions \cite{2019ApJ...886..133I}. 
Integral field unit (IFU) spectroscopy further enables direct mapping of the 
line-of-sight (LOS) velocity fields of the galaxies and provides robust measurements of 
their spins. 
Using spiral galaxy spins inferred from spiral-arm parities and IFU data, earlier
work reported a tentative correlation with the reconstructed primordial
tidal field close to the $3\sigma$ level \cite{2021NatAs...5..283M}.
However, a high-significance observational detection that directly links present-day galaxy spins 
to the cosmic initial conditions has remained lacking. 

In this Article, we present strong observational evidence for TTT by 
directly comparing observed galaxy spins with the reconstructed primordial tidal field. 
We measure the spins of both spiral and elliptical galaxies using IFU data 
from the latest release of the Mapping Nearby Galaxies at Apache Point Observatory 
(MaNGA) survey \cite{2022ApJS..259...35A}, and derive the primordial tidal 
field using the reconstructed primordial density perturbations from the ELUCID project 
\cite{2014ApJ...794...94W,2016ApJ...831..164W}. 
A direct observational implementation of the traditional TTT is not trivial, 
because its original formulation requires knowledge of protohalo quantities,  
which cannot be inferred directly from first principles in real observations.
To make such a comparison observationally possible, we apply the spin reconstruction framework 
\cite{2020PhRvL.124j1302Y} to observational data, 
using it to extract primordial angular momentum information from the reconstructed primordial 
tidal field.
Beyond establishing a high-significance observational detection, we also test the relation 
predicted by TTT between halo mass and the characteristic tidal scale from which angular 
momentum is generated, 
thereby linking the primordial tidal environment to galaxy-scale 
angular momentum acquisition in a directly testable way. 

\begin{figure}
	\centering
	 \includegraphics[width=1\linewidth]{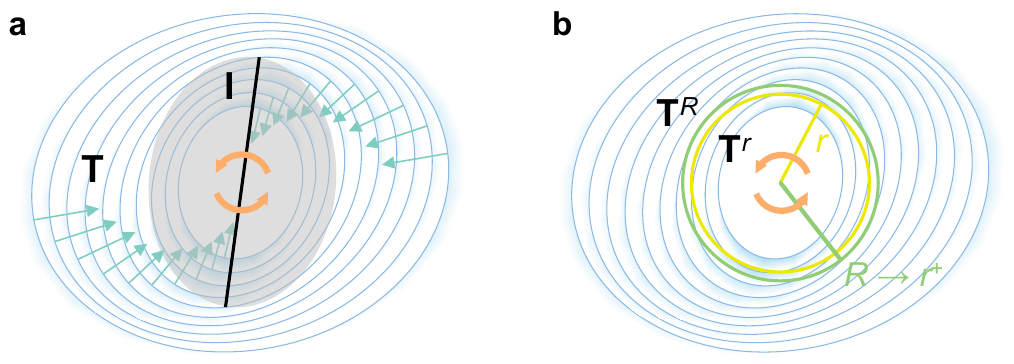}
	 \caption{\textbf{TTT and its spin reconstruction proxy.} 
	 \textbf{a}, Schematic illustration of the TTT. 
	 The grey filled ellipse marks the protohalo, whose major axis is shown by the black line.
	 Blue ellipses represent tidal tensors $\mathbf{T}$ smoothed on different scales.  
	 Cyan arrows indicate the rotation of their principal axes with increasing scale. 
	 Angular momentum arises from the misalignment between the protohalo inertia tensor 
	 $\mathbf{I}$ and the surrounding tidal tensor $\mathbf{T}$. 
	 \textbf{b}, Spin reconstruction used as a proxy for TTT. 
	 Concentric yellow and green circles indicate two smoothing scales, $r$ and $R$, which 
	 generate tidal tensors $\mathbf{T}^r$ and $\mathbf{T}^R$. 
	 The interaction between tidal fields smoothed on neighboring scales that bracket the 
	 protohalo size provides an approximation to its initial spin.
	 }
	 \label{fig.1}
\end{figure}

\section{Results}\label{sec2}
In TTT, the initial angular momentum of a protohalo is approximated as  
\begin{equation}
{\bs j}_{\rm TTT}=(j_\alpha)_{\rm TTT} \propto \epsilon_{\alpha\beta\gamma} 
I_{\beta\kappa} T_{\kappa\gamma}, 
	\label{eq:ttt}
\end{equation}
where ${\mathbf I}=(I_{\beta\kappa})$ is the protohalo inertia tensor and 
${\mathbf T}=(T_{\kappa\gamma})$ is the initial tidal tensor. 
The angular momentum arises from the misalignment between these 
two tensors, which is extracted by the Levi-Civita symbol $\epsilon_{\alpha\beta\gamma}$. 
Although the primordial tidal field can be obtained from a reconstruction of the cosmic 
initial condition, the protohalo shape is not analytically accessible from first principles. 
However, the protohalo inertia tensor is expected to be statistically linked to the 
anisotropic matter distribution around the halo, which is encoded in 
the reconstructed primordial tidal field. 
This motivates the use of the tidal field smoothed on different scales as a proxy for the 
inertia--tidal misalignment in equation~(\ref{eq:ttt}).
We, therefore, adopt a physically motivated proxy, known as spin 
reconstruction \cite{2020PhRvL.124j1302Y}, 
\begin{equation}
{\bs j}_{\rm R}=(j_\alpha)_{\rm R} \propto \epsilon_{\alpha\beta\gamma} T_{\beta\kappa}^r 
T_{\kappa\gamma}^R,
\label{eq:ttta}
\end{equation}
where ${\mathbf T}^r$ and ${\mathbf T}^R$ are the tidal fields smoothed at two scales, 
$r$ and $R$. 
Here, the smaller smoothing scale traces the protohalo-scale anisotropy, and the larger 
scale characterizes the surrounding tidal environment. 
Simulations show that this formulation approximates protohalo spin well when
(1) $r$ is close to $0.8 r_q$, with 
$r_{q}\equiv( {2M_{\rm h}G}/{\Omega_{\rm m} H_0^2})^{1/3},$
defining the Lagrangian radius of the DM halo, where $M_{\rm h}$, $G$, $\Omega_{\rm m}$, 
and $H_0$ denote the halo mass, Newton's constant, the matter density parameter and Hubble's 
constant, respectively, and 
(2) $R\rightarrow r^+$, where $r^+$ denotes the limit in which $R$ approaches $r$ from above. 
Figure~1 presents a schematic illustration of the TTT and the spin reconstruction 
framework. 

\begin{figure}
	\centering
	 \includegraphics[width=1\linewidth]{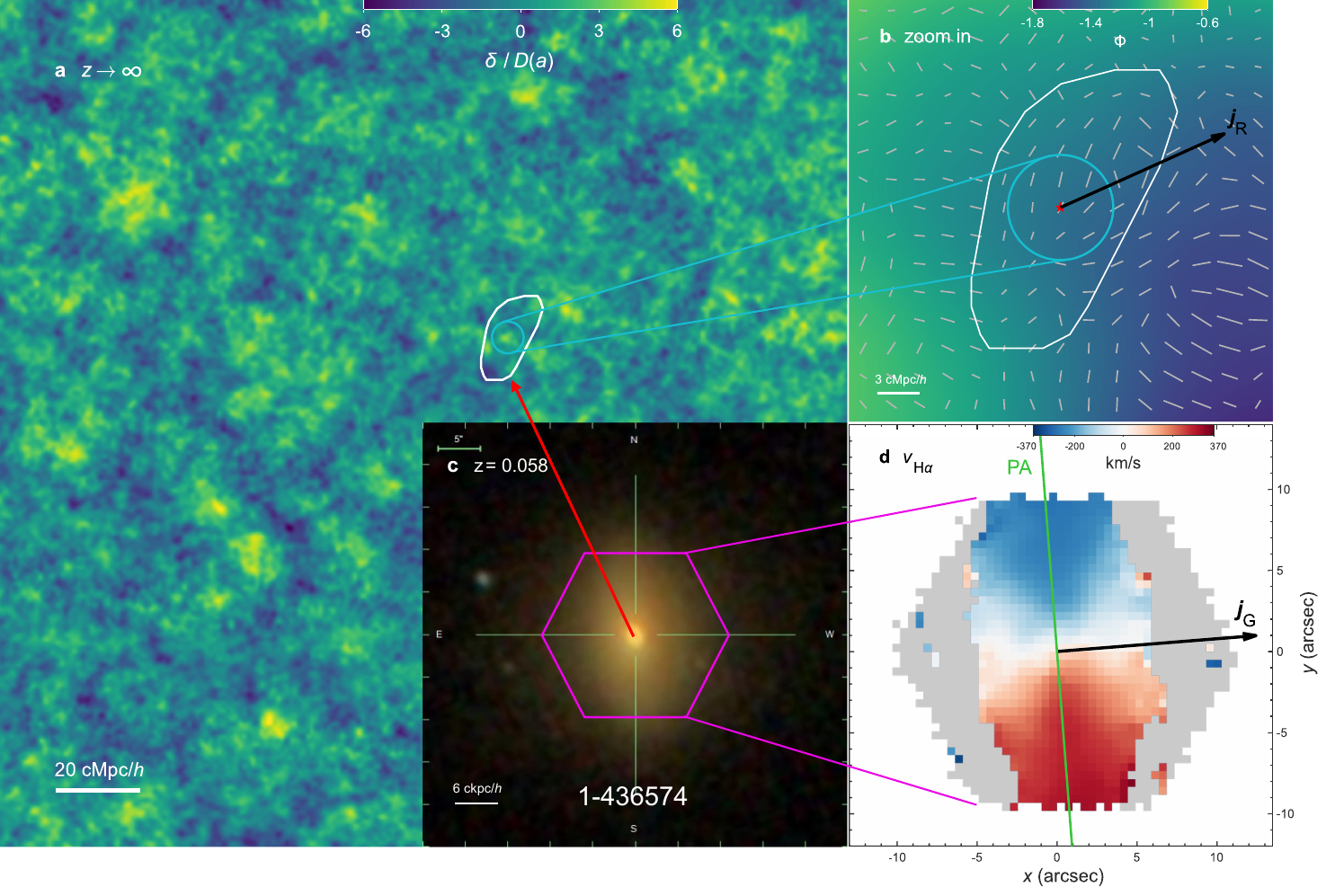}
	 \caption{\textbf{Observed and predicted spins for a representative galaxy.} 
	 \textbf{a}-\textbf{d}, A central elliptical galaxy from MaNGA (ID: 1-436574) with stellar 
	 mass $M_*=10^{11.3} M_{\odot}$ and halo mass $M_{\rm h}=10^{13.6} M_{\odot}$. 
	 \textbf{a}, Reconstructed primordial overdensity field of the nearby Universe from ELUCID,  
	 rescaled by linear growth factor as $\delta/D(a)$, where $\delta$ is the matter overdensity, 
	 $D(a)$ is the linear growth factor and $a$ is the cosmic scale factor, shown as a 
	 $200 \times 200 \times 10~{\rm cMpc}/h$ slice projected along the LOS of the observed galaxies. 
	 Here, $\rm cMpc$ and $\rm ckpc$ denote comoving megaparsecs and comoving kiloparsecs, respectively. 
	 \textbf{b}, Zoom-in on the protohalo. 
	 The background shows the primordial gravitational potential $\Phi$, smoothed on 
	 the scale $0.8 r_q$ and normalized by its standard deviation.
	 The short gray segments indicate the major axes of the tidal tensor derived from the 
	 smoothed potential at the corresponding locations, with their lengths rescaled according to the 
	 associated eigenvalues. The major axis marks the direction of strongest compression.
	 The red star marks the center of mass, and the white contour outlines the convex hull of 
	 the protohalo. 
	 The black arrow shows the predicted spin calculated within the $0.8 r_q$ 
	 region (blue circle). 
	 \textbf{c}, SDSS g-band, r-band, i-band composite image 
	 with the MaNGA bundle footprint (hexagon). The galaxy is at redshift $z=0.058$.
	 \textbf{d}, Gas velocity field traced by ${\rm H}\alpha$. The red and blue colours 
	 represent receding and approaching gas, respectively. 
	 Spaxels with ${\rm H}\alpha$ S/N $<$ 3 are masked in gray.
	 The green line marks the kinematic PA, and the black arrow shows the observed 
	 spin.
	 }
	 \label{fig.2}
\end{figure}

We obtained observational evidence for TTT by correlating the observed galaxy spins 
with the host protohalo spins predicted by equation~(\ref{eq:ttta}). 
The observed galaxy spin measurements were obtained from the MaNGA survey 
\cite{2015ApJ...798....7B}. 
We fitted the kinematic position angles (PAs) of gas (traced by ${\rm H}\alpha$) and stellar 
velocity fields using the Python-based package \texttt{PAFIT} \cite{2006MNRAS.366..787K}. 
The two-dimensional (2D) projected spins $\bs{j}_{\rm G}$, lying in the projected galaxy 
planes perpendicular to the LOS, were then derived separately for the gas and stellar 
components from their measured PAs and respective redshifted or blueshifted velocity 
distributions. 
To compute the predicted initial spins, we used the reconstructed primordial density field 
of the nearby Universe from the ELUCID project. 
The MaNGA galaxies were cross-matched with their host DM halos identified in refs.~
\cite{2005MNRAS.356.1293Y,2007ApJ...671..153Y}. 
We restricted our analysis to central galaxies residing in halos with masses 
$M_{\rm h}\geq 10^{12}M_{\odot}$, as these were within the region where ELUCID 
provides initial condition. 
We then employed a Lagrangian space remapping technique \cite{2024PhRvD.110b3512L} 
to trace these halos back to their initial, Lagrangian positions of the corresponding 
protohalos (`Lagrangian space remapping' in Methods). 
The predicted initial spins were subsequently calculated at the protohalo positions 
using equation~(\ref{eq:ttta}).
For a fair comparison with observations, we projected the predicted spins onto the 
planes of the corresponding galaxies, yielding the 2D-projected spins $\bs{j}_{\rm R}$. 
In Fig.~2, we present a representative example of the observed and predicted spins 
for a central elliptical galaxy. 

\begin{figure}
	\centering
	 \includegraphics[width=0.7\linewidth]{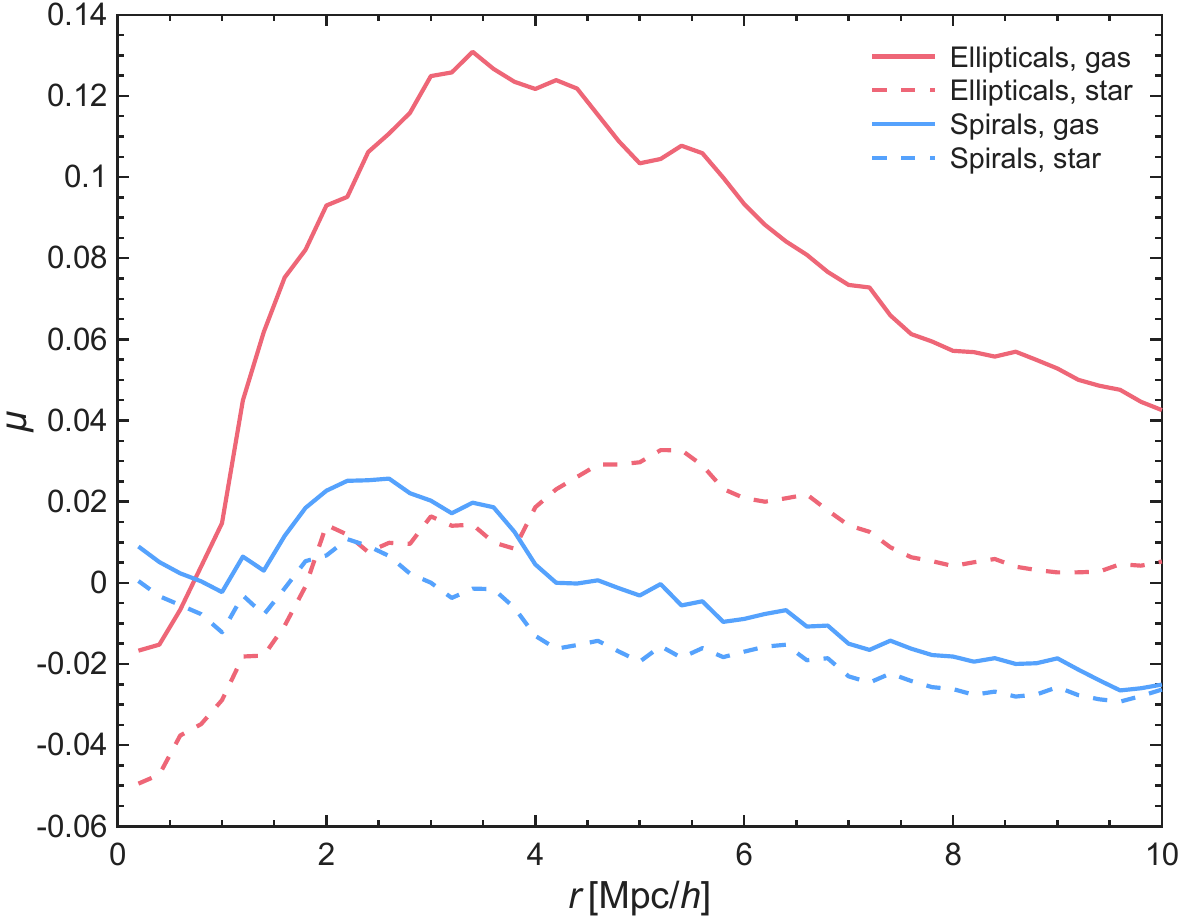}
	 \caption{\textbf{Spin--tidal field correlations as a function of smoothing scale.}
	 Red curves show central elliptical galaxies and blue curves show central spiral galaxies. 
	 Solid lines use gas kinematics as spin tracers; dashed lines use stellar kinematics. 
	 The larger smoothing scale $R$ is chosen to be slightly larger than $r$ ($R\rightarrow r^+$), 
	 following simulation results showing that this choice maximizes the correlation between the 
	 reconstructed and protohalo spins (`Results'). 
	 }	
	 \label{fig.3}
\end{figure}

By adopting the MaNGA galaxy morphological classifications \cite{2022MNRAS.509.4024D}, 
we obtained 781 central elliptical galaxies and 815 central spiral galaxies with reliable gas 
$\bs{j}_{\rm G}$ and 635 central elliptical galaxies and 812 central spiral galaxies with 
reliable stellar $\bs{j}_{\rm G}$.
For each galaxy, we define the direction correlation between $\bs{j}_{\rm G}$ and $\bs{j}_{\rm R}$ 
by the cosine of the angle between them: 
$\cos\theta_{\rm GR} = (\bs{j}_{\rm G}\cdot\bs{j}_{\rm R}) / |\bs{j}_{\rm G}||\bs{j}_{\rm R}| \in [-1,1]$.
In 2D space, randomly oriented (uncorrelated) vectors follow an arcsine distribution in 
$\cos\theta_{\rm GR}$, with expectation value $\left\langle \cos\theta_{\rm GR} \right\rangle= 0$.
In addition to the direction, the reconstructed primordial tidal field also contains information 
relevant to the angular momentum magnitude of protohalos. 
In particular, the tidal environment parameter $\beta_{\rm TT}$ quantifies the local efficiency of 
primordial tidal-torque generation on the protohalo scale \cite{2021PhRvD.103f3522W} 
(`Angular momentum magnitude in Lagrangian space' in Methods). 
Simulations show that protohalos with larger $\beta_{\rm TT}$ have more reliably predicted spins 
$\bs{j}_{\rm R}$ (ref.~\cite{2021PhRvD.103f3522W}). 
Motivated by this trend, we weighted each galaxy by $\beta_{\rm TT}^{\alpha}$, 
where $\alpha$ controls the sensitivity of this environmental weighting. 
Following the approximately square-root dependence found in simulations, we conservatively 
adopt $\alpha=1/2$. 
Varying $\alpha$ in the range $1/2\leq\alpha\leq2$ changed only the amplitude of the weighted 
correlation without changing its statistical significance or the main conclusions of this work.
The overall correlation of a galaxy sample is quantified by the weighted mean of \(\cos\theta_{\rm GR}\):
\begin{equation}
	\mu \equiv \frac{\sum_{i=1}^{N_{\rm galaxy}} \cos\theta_{{\rm GR},i}\,\beta_{{\rm TT},i}^{1/2}}
	{\sum_{i=1}^{N_{\rm galaxy}} \beta_{{\rm TT},i}^{1/2}}, 
\end{equation}
where $N_{\rm galaxy}$ is the number of galaxies in the sample.
The protohalo tidal twist distributions associated with the elliptical and spiral samples are broadly 
similar (Extended Data Fig. 1), so this weighting did not introduce a strong differential bias between
the two populations. 
Instead, it suppressed the contribution from halos with weak primordial tidal twists, whose spins are 
expected to be more weakly constrained and, therefore, more susceptible to random contamination. 
In our data, this weighting indeed led to a stronger spin--tidal field correlation signal.

Figure~3 shows the correlations as a function of universal smoothing scale for 
central elliptical and spiral galaxies, considering both gas and stellar components.
Gas in central elliptical galaxies exhibits the strongest correlation, reaching $\mu=0.13$ at the
optimal smoothing scale $r_{\rm opt}=3.4~{\rm Mpc}/h$ that is, the scale that maximizes the 
correlation signal. 
By contrast, central spiral galaxies exhibit much weaker correlations, and stellar spins are 
generally less well correlated than gas spins across both morphologies.
These results inidcate that the gas in central elliptical galaxies seems to be a better tracer 
of primordial tidal torques than the stellar component.	
Hereafter we restrict our analysis to the gas component of central elliptical galaxies. 

\begin{figure}
	\centering
	 \includegraphics[width=1\linewidth]{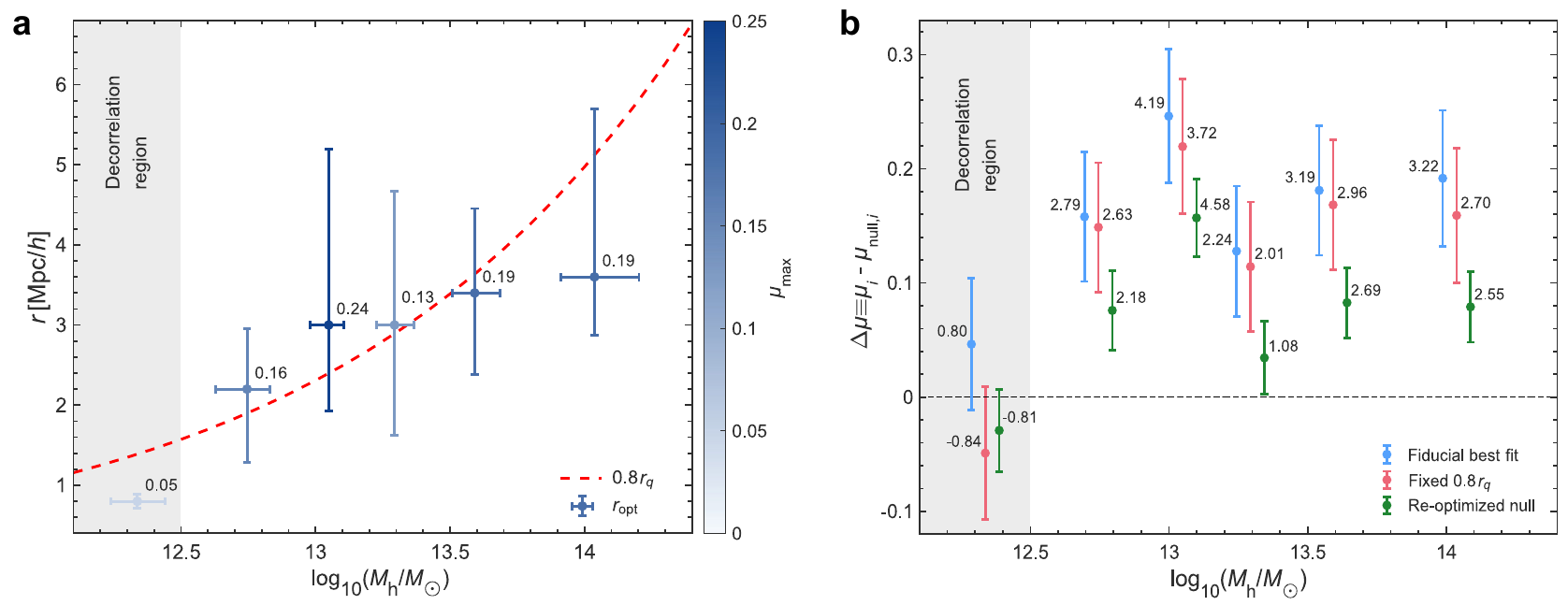}
	 \caption{\textbf{Mass dependence of the tidal scale of angular momentum generation and 
	 statistical significance of the spin--tidal field correlations.} 
	 \textbf{a}, Optimal smoothing scale as a function of halo mass. 
	 Each point shows the smoothing scale that maximizes the spin correlation $\mu_{\rm max}$ 
	 for galaxies in a given halo mass bin. 
	 Colours indicate the corresponding maximum correlation value, which is also labelled.
	 Vertical error bars mark the range of smoothing scales producing $\geq 68\%$ of the peak correlation, 
	 and horizontal error bars show the 25th to 75th percentiles of the halo mass distribution.
	 The red dashed curve shows the theoretically predicted optimal smoothing scale $0.8 r_q$, 
	 as previously found in numerical simulations. 
	 \textbf{b}, Statistical significance of the signal in each halo mass bin.
	 The points show the excess correlation relative to the shuffled null distributions, 
	 $\Delta\mu \equiv \mu_i-\mu_{\rm{null},\it i}$, 
	 and the error bars indicate the standard deviation of the corresponding null distribution.
	 Blue, red, and green denote the fiducial best fit, fixed scale, and re-optimized null cases, 
	 respectively, as defined in the main text.
	 The numbers label the significance of each mass bin, 
	 and the horizontal dashed line marks zero excess correlation.
	 The gray-shaded regions mark the `decorrelation region' in which the correlation 
	 and significance are suppressed owing to increasing Lagrangian space remapping errors and 
	 reconstruction uncertainties.}
	 \label{fig.4}
\end{figure}

Using the halo mass dependence of the optimal smoothing scale can further improve the correlation. 
We divided halos into mass bins that contained approximately equal numbers of samples. 
Figure~4a shows that, for most halo mass bins, $M_{\rm h}>10^{12.5}\,M_\odot$, the optimal smoothing 
scale increased systematically with halo mass, consistent with the predictions of TTT and numerical simulations, 
namely $r_{\rm opt} \approx 0.8 r_q$. 
In each mass bin, the search for the optimal smoothing scale was restricted to the interval 
$0.5$--$1.1 r_q$ to avoid unphysical peak selection at excessively small or large scales.
For the lowest mass bin, however, Lagrangian space remapping errors and reconstruction uncertainties 
became significant, leading to a suppression of the spin correlation 
(the shaded low-mass regime in Fig.~4a; see the analysis in refs.~\cite{2024PhRvD.110b3512L,2016ApJ...831..164W}).

To quantify the statistical significance of the detected correlation, we compared our measurements with 
null distributions constructed by randomly shuffling the observed spins within each halo mass bin. 
The null distribution construction and significance estimators are described in `Statistical analysis' in Methods.
Figure~4b shows the excess correlation $\Delta\mu \equiv \mu_i-\mu_{{\rm null},i}$ 
for three different prescriptions for defining the measured and null correlations. 
The lowest mass bin exhibited little or no excess signal and can even fell below the null expectation, 
consistent with the significant Lagrangian space remapping errors and reconstruction uncertainties expected 
in this regime. 
In the following significance analysis, we, therefore, excluded this lowest mass bin. 
Including it would reduce the overall significance, but would not affect the main conclusion. 

In the fiducial case, the measured correlation in each mass bin was evaluated at the optimal smoothing 
scale selected from the interval $0.5$--$1.1 r_q$, whereas the shuffled realizations were evaluated 
at that same fixed scale. Using the summary statistic $Z$ defined in 
`Statistical analysis' in Methods, the observed signal departed from the shuffled null expectation at the 
$7.0\sigma$ level. 
For comparison, we also considered two more stringent tests. 
In the fixed-scale case, both the measured and shuffled correlations were evaluated at the 
simulation-motivated scale $0.8 r_q$, yielding $Z=6.3\sigma$. 
In the re-optimized null case, the measured correlation was defined as in the fiducial analysis, 
but the optimal smoothing scales were selected for each shuffled realization independently 
over the same interval $0.5$--$1.1 r_q$, yielding $Z=5.9\sigma$. 
These tests were, therefore, either more a priori or more stringent than the fiducial analysis, 
yet both remained highly significant.

Taken together, these results provide robust observational evidence that present-day galaxy 
spins retain a measurable memory of primordial tidal torques, in support of TTT.

\section{Discussion}\label{sec3}
Among the galaxy populations considered here, the gas component of elliptical galaxies exhibits 
the strongest spin--tidal field correlation signal. 
This result probably reflects several effects acting in the same direction. 
First, the host halo mass distribution of the elliptical sample is more shifted toward higher 
masses than that of the spiral sample (Extended Data Fig.~2). 
This mass difference affects the signal in two ways: massive halos intrinsically retain 
a stronger connection to the primordial tidal field \cite{2023ApJ...943..128S}, and they occupy 
larger Lagrangian scales for which the ELUCID reconstruction \cite{2016ApJ...831..164W} and Lagrangian 
space remapping are more accurate \cite{2024PhRvD.110b3512L}. 
The lower typical host halo masses of spirals, therefore, naturally contribute to their weaker 
reconstructed spin--tidal field signal. 
Consistent with this interpretation, spirals exhibit detectable correlations once the lower mass 
systems are excluded (Extended Data Fig.~3).
However, central ellipticals still show stronger correlations at fixed halo mass, indicating that 
halo mass is not the sole driver of the observed difference. 
Recent observations indicate that the spin alignment between central galaxies and their host 
halos depends on morphology, with stronger alignment reported for ellipticals 
\cite{2025ApJ...992L..17W}, which may contribute to the residual difference. 
Second, the gas component may retain more memory of the primordial tidal field than the stellar 
component, as indicated by previous simulation-based work \cite{2023ApJ...943..128S}. 
Finally, in our elliptical sample, the gas component shows systematically smaller PA-fitting 
uncertainties than the stellar component (Extended Data Fig.~4), indicating that the ionized-gas 
velocity field is more kinematically coherent and that its spin can be measured more robustly. 
Taken together, these effects indicate that, within the present observational sample and 
reconstruction accuracy, the gas component of central ellipticals provides a better tracer 
of the early-time tidal-torque imprint preserved in present-day galaxy spins. 

This interpretation also highlights the central role of DM halos in connecting observable 
galaxies to their primordial environments.
Both observations and simulations indicate that the spins of central galaxies are closely correlated
with those of their host halos \cite{2025ApJ...992L..17W,2010MNRAS.404.1137B}, 
motivating the use of host halos as the bridge between observed galaxy spins and the primordial tidal field. 
In principle, galaxies can also be traced back to the initial conditions using the same inverse 
displacement field, and previous work has reported a spin--initial condition correlation through such a 
galaxy-based method \cite{2021NatAs...5..283M}.
However, the halos provide a more appropriate object for the Lagrangian space remapping in the present 
analysis because they correspond to larger physical and Lagrangian scales.
Numerical tests show that the relative remapping error increases toward lower mass systems 
\cite{2024PhRvD.110b3512L}, and the ELUCID-constrained simulation, being based 
on a halo-based group catalogue, is expected to reproduce halos more faithfully than individual galaxies. 
Consistent with this expectation, our tests show that directly remapping galaxies yielded a substantially 
weaker spin--tidal-field correlation than using their host halos. 
The halo-based remapping, therefore, provides a more reliable estimate of the relevant Lagrangian 
positions and contributes to the stronger signal recovered in the present analysis.

The remaining mismatch between observed galaxy spins and the spins predicted from the primordial 
tidal field reflects the combined impact of several stochastic and uncertain processes. 
Simulations indicate that the correlation decreases from about $0.7$ between protohalo and late-time 
halo spins \cite{2021PhRvD.103f3522W} to about $0.5$--$0.6$ when the protohalo spin 
is estimated by TTT \cite{2002MNRAS.332..325P} or the spin reconstruction proxy 
\cite{2020PhRvL.124j1302Y}, 
and to about $0.3$--$0.5$ for baryonic galaxy spins, with gas slightly more correlated than stars 
\cite{2023ApJ...943..128S}. 
These estimates do not include ELUCID reconstruction errors, Lagrangian space remapping 
uncertainties \cite{2024PhRvD.110b3512L},
observational PA errors, projection effects, halo mass and group assignment uncertainties. 
The signal measured here should, therefore, be regarded as a residual ensemble-level correlation, 
not a deterministic reconstruction of individual galaxy spins. 

The observed 3D spins of spiral galaxies can be derived by combining their kinematic PAs with the 
inclination angles estimated from the projected minor axis to major axis ratios \cite{2015ApJ...798...17Z}, 
under the assumption that both the gas and stellar components form a thin circular disk. 
For elliptical galaxies, this method is less straightforward: although in some elliptical galaxies 
the gas settles into a thin, rotationally supported disk --- making 3D gas spin estimates, in 
principle, feasible with IFU data --- their stellar component is typically triaxial and exhibits complex 
kinematics, which makes reliable 3D stellar spin measurements challenging. 
For consistency, therefore, we adopt 2D-projected spins for all galaxy types in this work. 
Although 2D projections remove one degree of directional information, they do not erase the intrinsic 
spin--tidal field correlation encoded in the full 3D directions.
Future high-resolution IFU observations and improved dynamical modeling may enable reliable 3D spin 
measurements for a broader range of galaxies. 
We leave these considerations for future work.

Our current IFU-based analysis yielded a final sample of 781 elliptical galaxies, but forthcoming 
large-scale spectroscopic surveys, such as by the Dark Energy Spectroscopic Instrument 
\cite{2013arXiv1308.0847L}, will observe millions of brightest cluster galaxies.
Although the fibres of the Dark Energy Spectroscopic Instrument provide only sparse spatial 
sampling, even a few (2-4) well-separated fibers per brightest cluster galaxy are sufficient 
to robustly determine their spins.
Moreover, existing reconstruction techniques of the cosmic initial condition --- such as ELUCID --- 
primarily rely on galaxy positions and have reached a bottleneck on small scales. 
Although small scales encode rich cosmological information and are crucial for addressing a wide range 
of cosmological questions, position- and density-based reconstruction methods have struggled for more 
than a decade to overcome this limitation.
One possible way to break this small-scale barrier is to incorporate other information, 
such as galaxy spins, which could be enabled by the large samples provided by next-generation 
surveys. 
At the current stage, the main observational goal is to establish a statistically significant 
detection of the spin--tidal field correlation in direction. 
Extending this connection to the angular momentum magnitude will require more precise measurements 
and a more complete theoretical understanding of the evolution of angular momentum in different galaxy 
components (for example, ref.~\cite{2024PhRvD.109l3548S}). 
Improved reconstruction accuracy and larger IFU samples will also be essential for extending the present 
detection from massive central ellipticals to lower mass haloes and broader galaxy populations, including 
spirals. 

The observational confirmation of TTT presented here therefore lays the groundwork for integrating 
angular momentum into future observation and reconstruction frameworks, 
which may open new dimensions for investigating the origin of the Universe  
and providing promising avenues for improving cosmological constraints and deepening our understanding of 
galaxy evolution.

\section{Methods}\label{sec4}	
\subsection{ELUCID reconstruction}\label{sec4.1}
The reconstructed primordial overdensity field $\delta$ of the nearby Universe was 
taken from the ELUCID project \cite{2014ApJ...794...94W,2016ApJ...831..164W}. 
The reconstruction started from a halo-based group catalogue 
\cite{2005MNRAS.356.1293Y,2007ApJ...671..153Y} built from Sloan Digital Sky Survey 
(SDSS) data, which we used to recover the present-day density field. 
In this catalogue, halo masses were estimated using the characteristic 
luminosity (or stellar mass) of group members and abundance matching. 
Comparisons with mock catalogues indicate a typical uncertainty of 
$\sim0.2$--$0.3$ dex in halo mass \cite{2007ApJ...671..153Y}.
Only galaxy groups in the Northern Galactic Cap with redshifts in the range 
$0.01\leq z \leq 0.12$ and luminosity-based group masses above $10^{12} M_\odot$ 
are included in the reconstruction. 
The `fingers of God' effect was largely suppressed during group construction, 
and the Kaiser effect was corrected in the reconstruction. 

The present-day density field was embedded in a cubic box of $500~{\rm Mpc}/h$ on each side and
discretized into $500^3$ cells with a spatial resolution of $1~{\rm Mpc}/h$. 
The initial density field was inferred from the present-day density field using a Hamiltonian 
Markov Chain Monte Carlo algorithm combined with particle-mesh dynamicswith 20 particle–mesh 
time steps \cite{2014ApJ...794...94W}. 
Evolving the reconstructed initial conditions to the present epoch with an $N$-body simulation 
reproduced the observed structures down to scales of $2~{\rm Mpc}/h$ (corresponding to halos 
with masses $M_{\rm h} \approx 10^{12.5} M_\odot$).

The primordial gravitational potential $\Phi$ was obtained from the reconstructed initial 
overdensity field by solving a normalized Poisson equation, $\nabla^2\Phi=\delta$, 
in Fourier space assuming periodic boundary conditions within the $500~{\rm Mpc}/h$ reconstruction volume. 
The potential was then smoothed with Gaussian kernels with scales $r$ and $R$. 
The tidal tensor $T_{ij}=-\partial_i\partial_j\Phi$ used in equation~(\ref{eq:ttta}) was subsequently 
computed in real space by evaluating the second derivatives of the smoothed potential using 
finite-difference operators on the grid.

\subsection{MaNGA galaxy sample}\label{sec4.2}
The MaNGA survey \cite{2015ApJ...798....7B}, 
part of the fourth-generation Sloan Digital Sky Survey (SDSS-IV) \cite{2017AJ....154...28B}, 
provides IFU spectroscopy and 2D maps of the LOS velocity fields for nearby galaxies. 
We derived from the final MaNGA data release \cite{2022ApJS..259...35A} the projected 2D spins 
of galaxies, which were defined to be perpendicular to the LOS.
The kinematic PAs of the gas and stellar components were measured separately using the \texttt{PAFIT} 
package \cite{2006MNRAS.366..787K}. 
The PA was defined as the anticlockwise angle from north to the line bisecting the velocity field. 
For the gas velocity field, we fitted only the spaxels with ${\rm H}\alpha$ signal-to-noise ratio 
S/N $\geq 3$, and for the stellar velocity field, we require a median spectral S/N $\geq 3$.
To ensure reliable PA measurements, we exclude `lineless' galaxies for which more than 90\% of 
spaxels within 1.5 effective radii $R_{\rm e}$ had ${\rm H}\alpha$ S/N $<$ 3, as well as systems with 
PA-fitting uncertainties $\phi_{\rm error} > 20^\circ$ (refs.~\cite{2022MNRAS.515.5081Z,2025ApJ...982L..29B}). 
The spin $\bs{j}_{\rm G}$ was oriented perpendicular to the kinematic PA, 
with its direction determined from the velocity field such that rotation proceeded from the 
receding (redshifted) side toward the approaching (blueshifted) side. 

MaNGA galaxies were cross-matched with the galaxy groups (host DM halos) \cite{2005MNRAS.356.1293Y,
2007ApJ...671..153Y} adopted in the ELUCID reconstruction. 
Following ELUCID, we used the luminosity-based group masses. The final sample was 
selected according to the following criteria:

(1) Central galaxies residing in groups (halos) with masses $M_{\rm h}\geq 10^{12}M_\odot$ 

(2) Galaxies within the reconstructed volume of ELUCID 

(3) Elliptical and spiral galaxies according to the classifications given in the 
MaNGA Deep Learning Morphology Value-Added Catalog \cite{2022MNRAS.509.4024D}. 

After applying the above selection and the PA-fitting quality cuts, we obtained 781 central 
elliptical galaxies and 815 central spiral galaxies with reliable gas spin measurements  
and 635 central elliptical galaxies and 812 central spiral galaxies with reliable stellar spin 
measurements. 
Because the PA-fitting quality cuts were imposed separately on the gas and stellar 
velocity fields, the corresponding samples did not fully overlap.

\subsection{Constrained simulation}\label{sec4.3}
Using the $500~{\rm Mpc}/h$ reconstructed initial condition provided by the ELUCID project, 
we ran a constrained cosmological $N$-body simulation with CUBE2 
\cite{2018ApJS..237...24Y,yu_cube2_2026}. 
Following ELUCID, we adopted the WMAP5 cosmology \cite{2009ApJS..180..306D} with parameters 
$\Omega_{\rm m}=0.258$, $\Omega_{\Lambda}=0.742$ (dark energy density parameter), 
$h=0.72$ (dimensionless Hubble parameter), $\sigma_8=0.796$ (linear matter fluctuation amplitude
on 8 ${\rm Mpc}/h$ scales), $n_s=0.963$ (primordial scalar spectral index). 
The simulation contained $N_{\rm p} = 500^3$ particles, initially placed uniformly in Lagrangian space 
using a `grid initial condition' so that each particle was at the center of a cell in 
a $N_{\rm g} = 500^3$ mesh. 
This set-up gave straightforward access to the Lagrangian properties of particles. 
The particles were then assigned linear displacements and peculiar velocities according to the 
Zel'dovich approximation \cite{1970A&A.....5...84Z} at an initial redshift of $z_{\rm init}=100$. 
Their subsequent evolution to Eulerian space at $z=0$ was computed using the 
Particle-Particle-Particle-Mesh gravity solver.

\subsection{Lagrangian space remapping}\label{sec4.4}
Lagrangian space remapping used the inverse displacement field from the constrained simulation 
to associate observed structures with estimated positions in the initial Lagrangian space 
\cite{2024PhRvD.110b3512L}. 
In the constrained simulation, the trajectories of $N$-body particles define a displacement 
field ${\bs \Psi}({\bs q})\equiv {\bs x}-{\bs q}$, which was uniformly sampled at the Lagrangian 
positions of the particles ${\bs q}$ and mapped them to their Eulerian positions ${\bs x}$. 
To remap structures back to the initial conditions, we constructed an inverse displacement field 
${\bs \Phi}({\bs x}) \equiv {\bs q}-{\bs x}$. 
Because this inverse field was not uniformly sampled by $N$-body particles in Eulerian space, 
it was estimated on the grid by averaging the displacement vectors of all particles that eventually 
occupy each grid cell. For an object with Eulerian centroid ${\bs x}$, the inverse displacement 
field ${\bs \Phi}$ defined on the grid was interpolated to ${\bs x}$ using the triangular-shaped 
cloud scheme \cite{1981csup.book.....H}. Its estimated Lagrangian position was then given by 
${\bs q}({\bs x}) = {\bs x} + {\bs \Phi}({\bs x})$.

\subsection{Angular momentum magnitude in Lagrangian space}\label{sec4.5}
The angular momentum magnitudes of Lagrangian protohalos can be characterized by the kinematic 
spin parameters \cite{2021PhRvD.103f3522W}. 
In particular, analogous to the spin reconstruction framework, 
a tidal environment parameter was defined directly from the reconstructed primordial tidal field as
\begin{equation}
\beta_{\rm TT} \equiv \frac{\left\lvert \epsilon_{\alpha\beta\gamma} 
T_{\beta\kappa}^r T_{\kappa\gamma}^R \right\rvert}
{T_{\alpha\beta}^r T_{\alpha\beta}^R}.
\end{equation}
Here, $\beta_{\rm TT}$ is a non-negative dimensionless quantity that characterizes 
the anisotropy and twist of ${\mathbf T}$ on physical scale $r$. It vanishes when no torque 
is generated and increases with the strength of the two-scale tidal misalignment.
We adopt $R\rightarrow r^+$ and $r=0.8 r_q$, 
so that $\beta_{\rm TT}$ characterizes the tidal environment on the protohalo scale.

\subsection{Statistical analysis}\label{sec4.6}
To assess the statistical significance of the detected spin--tidal field correlation, 
we generate 40,000 Monte Carlo realizations under the null hypothesis that the observed 
spins $\bs{j}_{\rm G}$ and the reconstructed spins $\bs{j}_{\rm R}$ are physically uncorrelated. 
In each realization, the observed spins were randomly permuted within each halo mass bin, 
thereby removing the galaxy-by-galaxy correspondence while preserving the intrinsic spin 
distribution, the halo mass distribution and the sample selection in each bin.
For consistency, the same \(\beta_{\rm TT}^{1/2}\) weighting was applied to the shuffled 
realizations when constructing the null distributions.

For each halo mass bin $i$, we denote the measured spin--tidal field correlation by $\mu_i$. 
For each prescription adopted in the main text, the shuffled realizations 
defined a corresponding null distribution for each mass bin. 
Denoting the mean and standard deviation of that null distribution by 
$\mu_{{\rm null},i}$ and $\sigma_{{\rm null},i}$, 
respectively, we defined a standardized significance for each mass bin as
\begin{equation}
z_i = \frac{\mu_i - \mu_{{\rm null},i}}{\sigma_{{\rm null},i}}.
\end{equation}
To characterize the overall departure from the null hypothesis across all mass bins, 
we defined a summary statistic: 
\begin{equation}
Z= \frac{\sum_{i=1}^{N_{\rm bin}} \left(\mu_i-\mu_{{\rm null},i}\right)}
{\left(\sum_{i=1}^{N_{\rm bin}} \sigma_{{\rm null},i}^2\right)^{1/2}}, 
\end{equation}
where $N_{\rm bin}$ is the number of bins.

Because each halo was in only one halo mass bin, the object samples in different bins were disjoint. 
This substantially reduced inter-bin covariance. However, as all bins were evaluated using the same 
reconstructed tidal field, a weak covariance between bins could not be completely excluded. 
We, therefore, regarded the summary significances quoted above as an approximate summary measures of the 
detection significance.

\section*{Data availability}
The data used for this paper, including the observed and reconstructed galaxy spins, are available via 
Zenodo at \url{https://doi.org/10.5281/zenodo.20263913} (ref.~\cite{sheng_2026_20263913}). 
The repository also includes a Python/Jupyter notebook to enable one to quickly reproduce the main 
diagnostic plots and check the spin correlation measurements.
The MaNGA data are publicly available at \url{https://www.sdss4.org/dr17/manga/manga-data/data-access/} 
and can be accessed via Marvin at \url{https://magrathea.sdss.org/marvin/}. 
The Morphology Value-added Catalogs are available at 
\url{https://data.sdss.org/sas/dr17/env/MANGA_MORPHOLOGY/deep_learning/}. 
The galaxy group catalogue is publicly available via the SDSS DR7 part of 
\url{https://gax.sjtu.edu.cn/data/Group.html}. 
The ELUCID reconstructed initial conditions are available from the corresponding author upon reasonable 
request.

\section*{Code availability}
The CUBE2 code has been publicly released and is available via GitHub at 
\url{https://github.com/yuhaoran/CUBE2} (ref.~\cite{2026ascl.soft04001Y}). 
The spin reconstruction, spin parameter calculation and Lagrangian remapping codes used 
in this study are publicly available via GitHub at \url{https://github.com/Hiki1998/TTT-observation}.

\section*{Acknowledgements}
We thank Qing Gu, Min Du and Peng Wang for valuable discussions and comments. 
We also acknowledge Manqi Fu, Wenyu Zhong, Hong-Chuan Ma, Dawei Li and Kun-Peng Shi for their help in visualization and 
data processing. 
We thank the reviewers for their valuable and constructive suggestions. 
We are especially grateful to Reviewer Mark Neyrinck for suggesting the use of angular momentum 
magnitude information, which helped further improve the final correlation signal and its statistical 
significance. 
We also sincerely thank the anonymous reviewer for the insightful comments that helped us better clarify 
the physical interpretation of the observed signal.

SDSS-IV acknowledges support and resources from the Center for High-Performance Computing at the University of Utah. 
The SDSS web site is \url{www.sdss.org}.
SDSS-IV is managed by the Astrophysical Research Consortium for the Participating Institutions of the SDSS Collaboration 
including the Brazilian Participation Group, the Carnegie Institution for Science, Carnegie Mellon University,
the Chilean Participation Group, the French Participation Group, Harvard-Smithsonian Center for Astrophysics, 
Instituto de Astrof\'{ı}sica de Canarias, The Johns Hopkins University, Kavli Institute for the Physics and Mathematics of
the Universe (IPMU)/University of Tokyo, Lawrence Berkeley National Laboratory, Leibniz Institut f\"{u}r Astrophysik
Potsdam (AIP), Max-Planck-Institut f\"{u}r Astronomie (MPIA Heidelberg), Max-Planck-Institut f\"{u}r Astrophysik (MPA
Garching), Max-Planck-Institut f\"{u}r Extraterrestrische Physik (MPE), National Astronomical Observatory of China,
New Mexico State University, New York University, University of Notre Dame, Observat\'{o}rio Nacional/MCTI, The
Ohio State University, Pennsylvania State University, Shanghai Astronomical Observatory, United Kingdom Participation 
Group, Universidad Nacional Aut\'{o}noma de M\'{e}xico, University of Arizona, University of Colorado Boulder,
University of Oxford, University of Portsmouth, University of Utah, University of Virginia, University of Washington,
University of Wisconsin, Vanderbilt University, and Yale University.

\section*{Funding}
M.-J.S. and H.-R.Y. acknowledge support from the National Natural Science Foundation of China (NSFC) 
grant No.~124B2054, No.~12173030 and the Fundamental Research Funds for the Central Universities 
No.~20720240149. 
The authors acknowledge the support by the China Manned Space Program through its Space Application
System. 
M.B. acknowledge support from the NSFC grant No.~12303009. 
Q.G. acknowledge support from the NSFC grant No.~12425303. 
Y.-M.C. acknowledges support from NSFC grant No.~12333002. 
H.W. is supported by the NSFC grant No.~12595312, 12192224 and CAS Project for Young Scientists in Basic Research, 
Grant No.~YSBR-062.
U.-L.P. is supported by the Natural Sciences and Engineering Research Council of Canada (NSERC) [funding reference number 
RGPIN-2019-06770, ALLRP 586559-23, RGPIN-2025-06396] , Canadian Institute for Advanced Research (CIFAR), Ontario Research 
Fund -- Research Excellence (ORF-RE Fund, 72074697), AMD AI Quantum Astro. 
X.Y. is supported by the National Key R\&D Program of China (2023YFA1607800, 2023YFA1607804) and the China Manned Space 
Project with No.~CMS-CSST-2025-A04. 
Funding for the Sloan Digital Sky Survey IV has been provided by the Alfred P. Sloan Foundation, the 
U.S. Department of Energy Office of Science, and the Participating Institutions.

\section*{Author contribution}
All authors contributed to the work presented in this paper. 
H.-R.Y, Q.G. and M.-J.S. initiated the project. 
M.-J.S. performed the data analysis and wrote the initial manuscript draft.
M.B. and Y.-M.C. contributed to the analysis of the observational data.
H.-R.Y and B.-H.C. developed the $N$-body simulation code. 
H.W. and X.Y. provided the ELUCID group catalogues and the reconstructed initial conditions.
Q.G., U.-L.P., and J.W. provided guidance on the theoretical interpretation and scientific 
significance of the results.
H.-R.Y., M.B., and F.-N.S. helped revise and refine the manuscript.
All authors participated in discussions that shaped the direction of the project and influenced 
the final manuscript.

\section*{Competing interests}
The authors declare no competing interests.

\clearpage

\renewcommand{\figurename}{\bf Extended Data Fig.}
\setcounter{figure}{0}

\begin{figure}
	\centering
	 \includegraphics[width=0.7\linewidth]{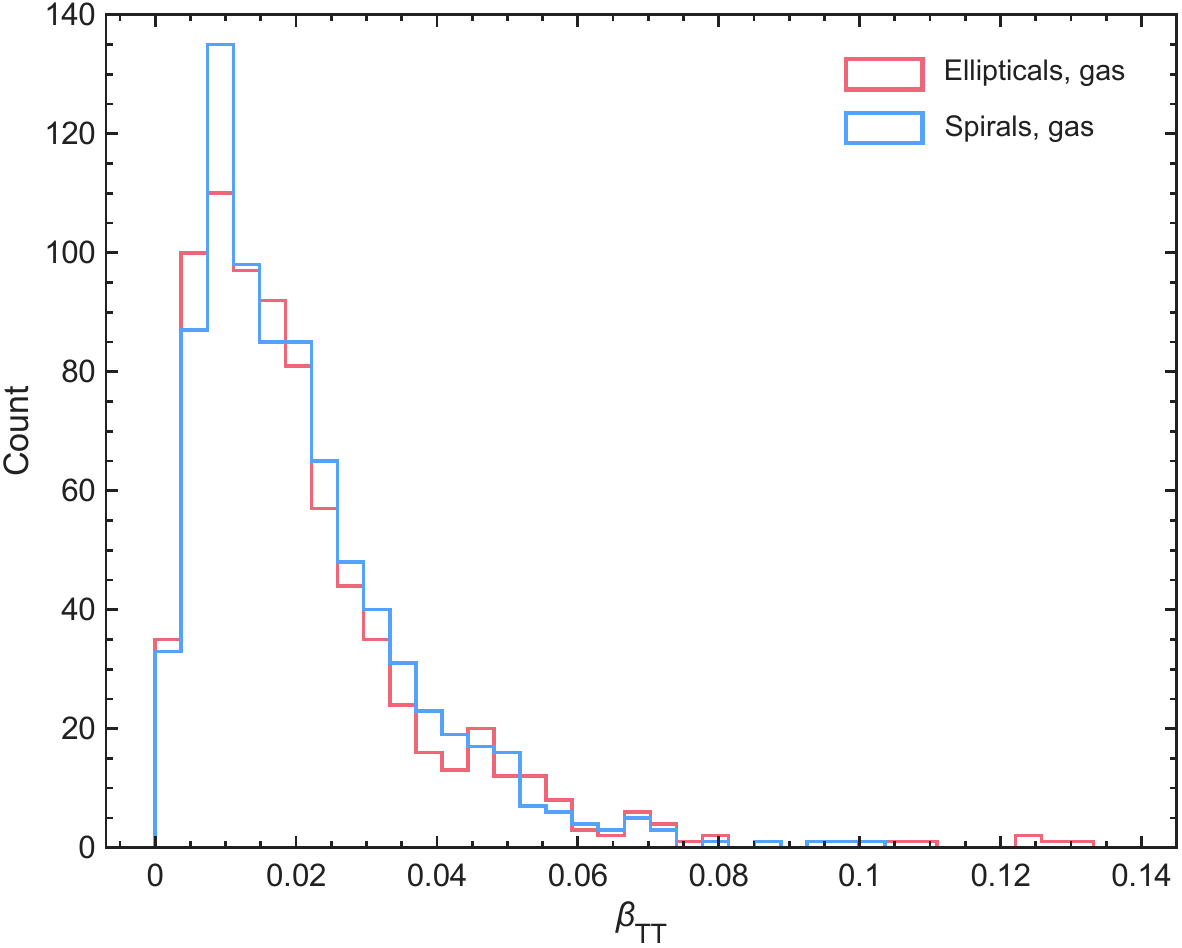}
	 \caption{\textbf{Distribution of protohalo tidal environment parameter $\beta_{\rm TT}$ for 
	 the elliptical and spiral galaxy samples.}
	 The histograms show the host halos of 781 elliptical galaxies and 815 spiral galaxies with 
	 reliable gas spin measurements used in the main analysis. The two samples exhibit broadly similar 
	 $\beta_{\rm TT}$ distributions, indicating that the stronger spin--tidal field correlation 
	 detected for ellipticals is not primarily driven by differences in the tidal environment parameter. 
	 }
	 \label{fig.ex1}
\end{figure}

\clearpage

\begin{figure}
	\centering
	 \includegraphics[width=0.7\linewidth]{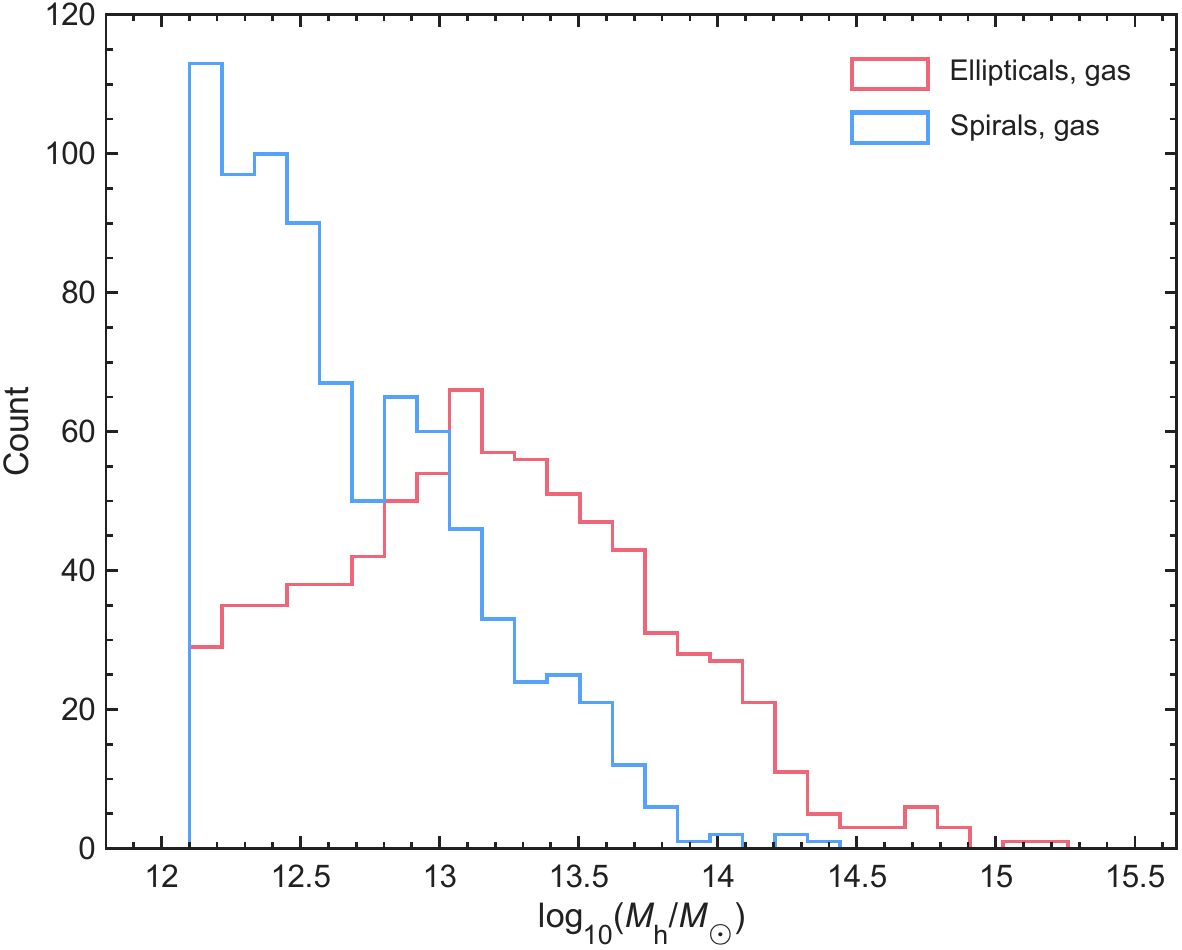}
	 \caption{\textbf{Distribution of host halo masses for the elliptical and spiral galaxy samples.}
	 The samples contain the host halos of 781 elliptical galaxies and 815 spiral galaxies with reliable 
	 gas spin measurements used in the main analysis. The spiral sample is systematically shifted toward 
	 lower halo masses than the elliptical sample. A substantial fraction of the spiral host halos lie 
	 below $10^{12.5}\,M_\odot$, i.e. in the `decorrelation region' where reconstruction uncertainties 
	 and Lagrangian space remapping errors are expected to be significant, thereby reducing the detected 
	 spin--tidal field correlation signal.
	 }	
	 \label{fig.ex2}
\end{figure}

\clearpage

\begin{figure}
	\centering
	 \includegraphics[width=1\linewidth]{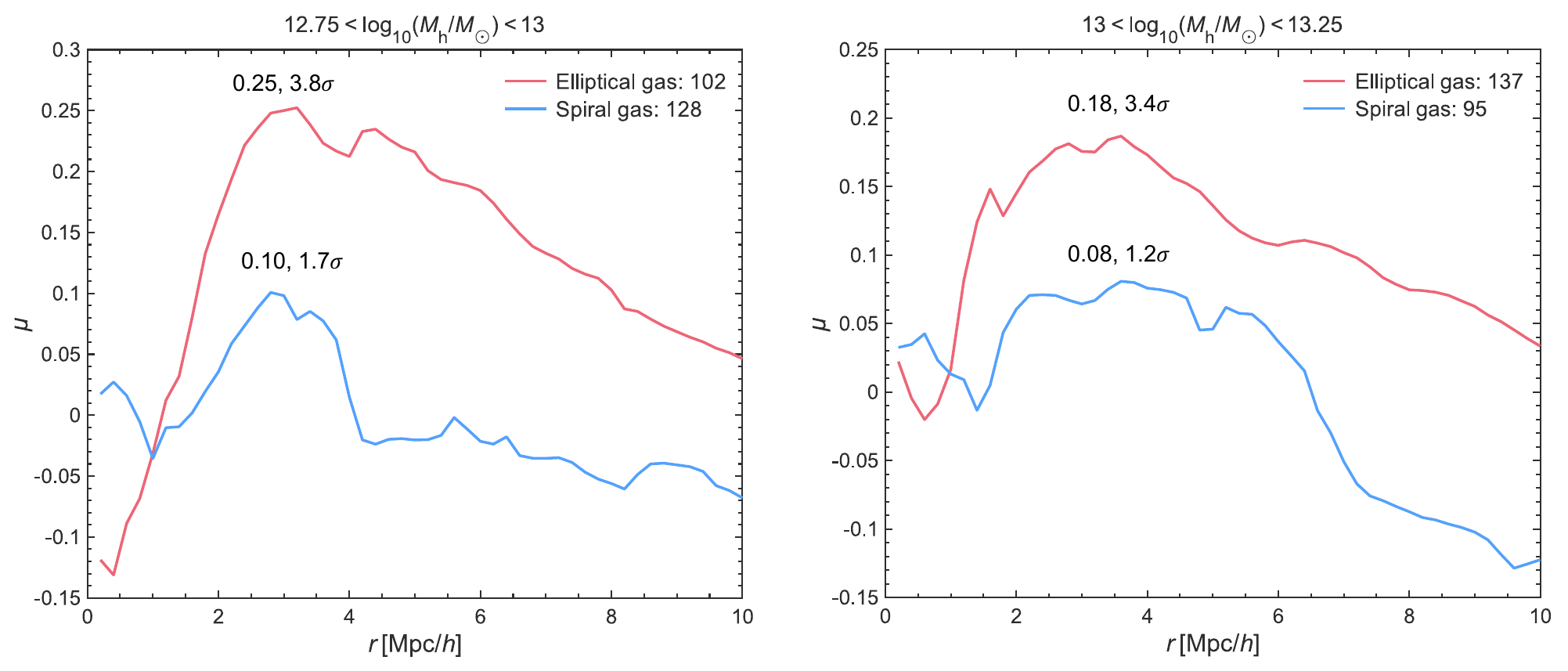}
	 \caption{\textbf{Mass-controlled comparison of central elliptical and spiral galaxies.}
	 Spin--tidal field correlations for gas in central ellipticals and spirals are shown in two restricted 
	 host-halo mass bins, $12.75<\log_{10}(M_{\rm h}/M_\odot)<13$ and $13<\log_{10}(M_{\rm h}/M_\odot)<13.25$. 
	 The sample size, mean correlation and detection significance are labelled in each panel. 
	 Spiral galaxies exhibit detectable correlations once lower-mass systems are excluded, indicating that 
	 the weak signal in the full spiral sample is partly driven by host-halo mass. At fixed halo mass, however, 
	 central ellipticals still show stronger correlations than spirals. 
	 }	
	 \label{fig.ex3}
\end{figure}

\clearpage

\begin{figure}
	\centering
	 \includegraphics[width=0.7\linewidth]{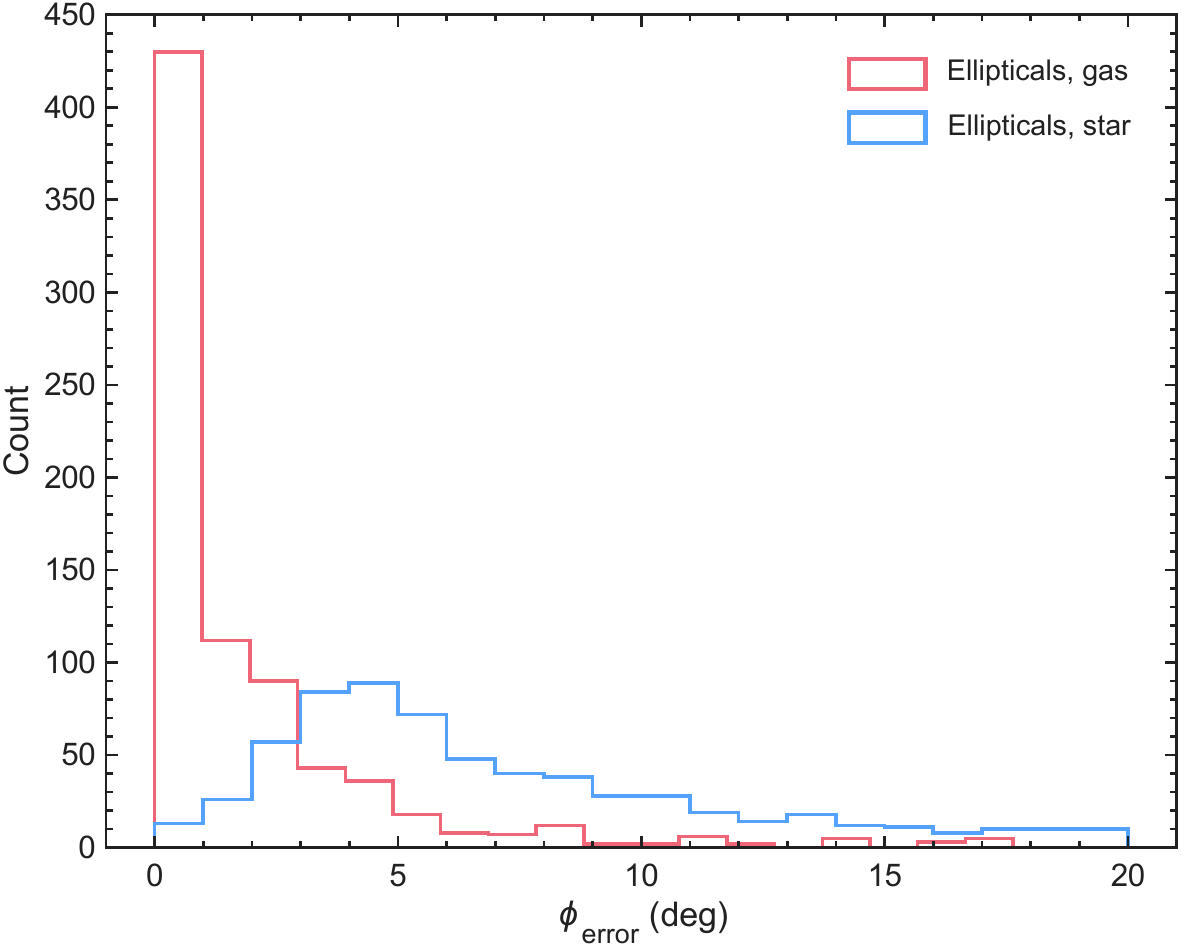}
	 \caption{\textbf{Distribution of PA fitting uncertainties $\phi_{\rm error}$ for the gas and stellar 
	 components of elliptical galaxies.} 
	 The gas and stellar distributions are each restricted to galaxies with $\phi_{\rm error}<20^\circ$ in 
	 the corresponding component. Smaller PA fitting uncertainties indicate more kinematically 
	 coherent velocity fields and therefore more robust spin measurements.
	 }
	 \label{fig.ex4}
\end{figure}


\clearpage

\begin{thebibliography}{10}
\expandafter\ifx\csname url\endcsname\relax
  \def\url#1{\burl{#1}}\fi
\expandafter\ifx\csname urlprefix\endcsname\relax\def\urlprefix{URL }\fi
\providecommand{\bibinfo}[2]{#2}
\providecommand{\eprint}[2][]{\url{#2}}
\providecommand{\doi}[1]{\url{https://doi.org/#1}}

\bibitem{1980MNRAS.193..189F}
\bibinfo{author}{{Fall}, S.~M.} \& \bibinfo{author}{{Efstathiou}, G.}
\newblock \bibinfo{title}{{Formation and rotation of disc galaxies with haloes.}}
\newblock \textit{\bibinfo{journal}{Mon. Not. R. Astron. Soc.}} \textbf{\bibinfo{volume}{193}}, \bibinfo{pages}{189--206} (\bibinfo{year}{1980}).

\bibitem{1998MNRAS.295..319M}
\bibinfo{author}{{Mo}, H.~J.}, \bibinfo{author}{{Mao}, S.} \& \bibinfo{author}{{White}, S. D.~M.}
\newblock \bibinfo{title}{{The formation of galactic discs}}.
\newblock \textit{\bibinfo{journal}{Mon. Not. R. Astron. Soc.}} \textbf{\bibinfo{volume}{295}}, \bibinfo{pages}{319--336} (\bibinfo{year}{1998}).

\bibitem{2014ApJ...784...26O}
\bibinfo{author}{{Obreschkow}, D.} \& \bibinfo{author}{{Glazebrook}, K.}
\newblock \bibinfo{title}{{Fundamental Mass-Spin-Morphology Relation Of Spiral Galaxies}}.
\newblock \textit{\bibinfo{journal}{Astrophys. J.}} \textbf{\bibinfo{volume}{784}}, \bibinfo{pages}{26} (\bibinfo{year}{2014}).

\bibitem{2024JCAP...05..111M}
\bibinfo{author}{{Moon}, J.-S.} \& \bibinfo{author}{{Lee}, J.}
\newblock \bibinfo{title}{{Mutual information between galaxy properties and the initial predisposition}}.
\newblock \textit{\bibinfo{journal}{J. Cosmol. Astropart. Phys.}} \textbf{\bibinfo{volume}{2024}}, \bibinfo{pages}{111} (\bibinfo{year}{2024}).

\bibitem{2005MNRAS.360L..82R}
\bibinfo{author}{{Rimes}, C.~D.} \& \bibinfo{author}{{Hamilton}, A. J.~S.}
\newblock \bibinfo{title}{{Information content of the non-linear matter power spectrum}}.
\newblock \textit{\bibinfo{journal}{Mon. Not. R. Astron. Soc.}} \textbf{\bibinfo{volume}{360}}, \bibinfo{pages}{L82--L86} (\bibinfo{year}{2005}).

\bibitem{2021JCAP...06..024M}
\bibinfo{author}{{McQuinn}, M.}
\newblock \bibinfo{title}{{On the primordial information available to galaxy redshift surveys}}.
\newblock \textit{\bibinfo{journal}{J. Cosmol. Astropart. Phys.}} \textbf{\bibinfo{volume}{2021}}, \bibinfo{pages}{024} (\bibinfo{year}{2021}).

\bibitem{2020ApJ...898L..27L}
\bibinfo{author}{{Lee}, J.}, \bibinfo{author}{{Libeskind}, N.~I.} \& \bibinfo{author}{{Ryu}, S.}
\newblock \bibinfo{title}{{The Effect of Massive Neutrinos on the Halo Spin Flip Phenomenon}}.
\newblock \textit{\bibinfo{journal}{Astrophys. J. Lett.}} \textbf{\bibinfo{volume}{898}}, \bibinfo{pages}{L27} (\bibinfo{year}{2020}).

\bibitem{2020PhRvL.124j1302Y}
\bibinfo{author}{{Yu}, H.-R. et~al.} 
\newblock \bibinfo{title}{{Probing Primordial Chirality with Galaxy Spins}}.
\newblock \textit{\bibinfo{journal}{Phys. Rev. Lett.}} \textbf{\bibinfo{volume}{124}}, \bibinfo{pages}{101302} (\bibinfo{year}{2020}).

\bibitem{2025PhRvL.135n1002S}
\bibinfo{author}{{Shim}, J.}, \bibinfo{author}{{Pen}, U.-L.}, \bibinfo{author}{{Yu}, H.-R.} \& \bibinfo{author}{{Okumura}, T.}
\newblock \bibinfo{title}{{Probing Vector Chirality in the Early Universe}}.
\newblock \textit{\bibinfo{journal}{Phys. Rev. Lett.}} \textbf{\bibinfo{volume}{135}}, \bibinfo{pages}{141002} (\bibinfo{year}{2025}).

\bibitem{2020ApJ...902...22L}
\bibinfo{author}{{Lee}, J.} \& \bibinfo{author}{{Libeskind}, N.~I.}
\newblock \bibinfo{title}{{The Halo Spin Transition As a Probe of Dark Energy}}.
\newblock \textit{\bibinfo{journal}{Astrophys. J.}} \textbf{\bibinfo{volume}{902}}, \bibinfo{pages}{22} (\bibinfo{year}{2020}).

\bibitem{2025arXiv251110005D}
\bibinfo{author}{{da Silveira Ferreira}, P. et~al.}
\newblock \bibinfo{title}{{Where Galaxies Point: First Measurement of the Large-Scale Axial Intrinsic Alignment}}.
\newblock \textit{\bibinfo{journal}{arXiv e-prints}} \bibinfo{pages}{arXiv:2511.10005} (\bibinfo{year}{2025}).

\bibitem{1969ApJ...155..393P}
\bibinfo{author}{{Peebles}, P.~J.~E.}
\newblock \bibinfo{title}{{Origin of the Angular Momentum of Galaxies}}.
\newblock \textit{\bibinfo{journal}{Astrophys. J.}} \textbf{\bibinfo{volume}{155}}, \bibinfo{pages}{393} (\bibinfo{year}{1969}).

\bibitem{1970Afz.....6..581D}
\bibinfo{author}{{Doroshkevich}, A.~G.}
\newblock \bibinfo{title}{{The space structure of perturbations and the origin of rotation of galaxies in the theory of fluctuation.}}
\newblock \textit{\bibinfo{journal}{Astrofizika}} \textbf{\bibinfo{volume}{6}}, \bibinfo{pages}{581--600} (\bibinfo{year}{1970}).

\bibitem{1984ApJ...286...38W}
\bibinfo{author}{{White}, S.~D.~M.}
\newblock \bibinfo{title}{{Angular momentum growth in protogalaxies}}.
\newblock \textit{\bibinfo{journal}{Astrophys. J.}} \textbf{\bibinfo{volume}{286}}, \bibinfo{pages}{38--41} (\bibinfo{year}{1984}).

\bibitem{2002MNRAS.332..325P}
\bibinfo{author}{{Porciani}, C.}, \bibinfo{author}{{Dekel}, A.} \& \bibinfo{author}{{Hoffman}, Y.}
\newblock \bibinfo{title}{{Testing tidal-torque theory - I. Spin amplitude and direction}}.
\newblock \textit{\bibinfo{journal}{Mon. Not. R. Astron. Soc.}} \textbf{\bibinfo{volume}{332}}, \bibinfo{pages}{325--338} (\bibinfo{year}{2002}).

\bibitem{2000ApJ...532L...5L}
\bibinfo{author}{{Lee}, J.} \& \bibinfo{author}{{Pen}, U.-L.}
\newblock \bibinfo{title}{{Cosmic Shear from Galaxy Spins}}.
\newblock \textit{\bibinfo{journal}{Astrophys. J. Lett.}} \textbf{\bibinfo{volume}{532}}, \bibinfo{pages}{L5--L8} (\bibinfo{year}{2000}).

\bibitem{2001ApJ...555..106L}
\bibinfo{author}{{Lee}, J.} \& \bibinfo{author}{{Pen}, U.-L.}
\newblock \bibinfo{title}{{Galaxy Spin Statistics and Spin-Density Correlation}}.
\newblock \textit{\bibinfo{journal}{Astrophys. J.}} \textbf{\bibinfo{volume}{555}}, \bibinfo{pages}{106--124} (\bibinfo{year}{2001}).

\bibitem{2021PhRvD.103f3522W}
\bibinfo{author}{{Wu}, Q.}, \bibinfo{author}{{Yu}, H.-R.}, \bibinfo{author}{{Liao}, S.} \& \bibinfo{author}{{Du}, M.}
\newblock \bibinfo{title}{{Spin mode reconstruction in Lagrangian space}}.
\newblock \textit{\bibinfo{journal}{Phys. Rev. D}} \textbf{\bibinfo{volume}{103}}, \bibinfo{pages}{063522} (\bibinfo{year}{2021}).

\bibitem{2010MNRAS.404.1137B}
\bibinfo{author}{{Bett}, P.}, \bibinfo{author}{{Eke}, V.}, \bibinfo{author}{{Frenk}, C.~S.}, \bibinfo{author}{{Jenkins}, A.} \& \bibinfo{author}{{Okamoto}, T.}
\newblock \bibinfo{title}{{The angular momentum of cold dark matter haloes with and without baryons}}.
\newblock \textit{\bibinfo{journal}{Mon. Not. R. Astron. Soc.}} \textbf{\bibinfo{volume}{404}}, \bibinfo{pages}{1137--1156} (\bibinfo{year}{2010}).

\bibitem{2023ApJ...943..128S}
\bibinfo{author}{{Sheng}, M.-J. et~al.} 
\newblock \bibinfo{title}{{Baryonic Effects on Lagrangian Clustering and Angular Momentum Reconstruction}}.
\newblock \textit{\bibinfo{journal}{Astrophys. J.}} \textbf{\bibinfo{volume}{943}}, \bibinfo{pages}{128} (\bibinfo{year}{2023}).

\bibitem{2024PhRvD.109l3548S}
\bibinfo{author}{{Sheng}, M.-J. et~al.}
\newblock \bibinfo{title}{{Spin speed correlations and the evolution of galaxy-halo systems}}.
\newblock \textit{\bibinfo{journal}{Phys. Rev. D}} \textbf{\bibinfo{volume}{109}}, \bibinfo{pages}{123548} (\bibinfo{year}{2024}).

\bibitem{2019ApJ...886..133I}
\bibinfo{author}{{Iye}, M.}, \bibinfo{author}{{Tadaki}, K.-i.} \& \bibinfo{author}{{Fukumoto}, H.}
\newblock \bibinfo{title}{{Spin Parity of Spiral Galaxies. I. Corroborative Evidence for Trailing Spirals}}.
\newblock \textit{\bibinfo{journal}{Astrophys. J.}} \textbf{\bibinfo{volume}{886}}, \bibinfo{pages}{133} (\bibinfo{year}{2019}).

\bibitem{2021NatAs...5..283M}
\bibinfo{author}{{Motloch}, P.}, \bibinfo{author}{{Yu}, H.-R.}, \bibinfo{author}{{Pen}, U.-L.} \& \bibinfo{author}{{Xie}, Y.}
\newblock \bibinfo{title}{{An observed correlation between galaxy spins and initial conditions}}.
\newblock \textit{\bibinfo{journal}{Nat. Astron.}} \textbf{\bibinfo{volume}{5}}, \bibinfo{pages}{283--288} (\bibinfo{year}{2021}).

\bibitem{2022ApJS..259...35A}
\bibinfo{author}{{Abdurro'uf} et~al.}
\newblock \bibinfo{title}{{The Seventeenth Data Release of the Sloan Digital Sky Surveys: Complete Release of MaNGA, MaStar, and APOGEE-2 Data}}.
\newblock \textit{\bibinfo{journal}{Astrophys. J. Suppl.}} \textbf{\bibinfo{volume}{259}}, \bibinfo{pages}{35} (\bibinfo{year}{2022}).

\bibitem{2014ApJ...794...94W}
\bibinfo{author}{{Wang}, H.}, \bibinfo{author}{{Mo}, H.~J.}, \bibinfo{author}{{Yang}, X.}, \bibinfo{author}{{Jing}, Y.~P.} \& \bibinfo{author}{{Lin}, W.~P.}
\newblock \bibinfo{title}{{ELUCID{\textemdash}Exploring the Local Universe with the Reconstructed Initial Density Field. I. Hamiltonian Markov Chain Monte Carlo Method with Particle Mesh Dynamics}}.
\newblock \textit{\bibinfo{journal}{Astrophys. J.}} \textbf{\bibinfo{volume}{794}}, \bibinfo{pages}{94} (\bibinfo{year}{2014}).

\bibitem{2016ApJ...831..164W}
\bibinfo{author}{{Wang}, H. et~al.}
\newblock \bibinfo{title}{{ELUCID - Exploring the Local Universe with ReConstructed Initial Density Field III: Constrained Simulation in the SDSS Volume}}.
\newblock \textit{\bibinfo{journal}{Astrophys. J.}} \textbf{\bibinfo{volume}{831}}, \bibinfo{pages}{164} (\bibinfo{year}{2016}).

\bibitem{2015ApJ...798....7B}
\bibinfo{author}{{Bundy}, K. et~al.}
\newblock \bibinfo{title}{{Overview of the SDSS-IV MaNGA Survey: Mapping nearby Galaxies at Apache Point Observatory}}.
\newblock \textit{\bibinfo{journal}{Astrophys. J.}} \textbf{\bibinfo{volume}{798}}, \bibinfo{pages}{7} (\bibinfo{year}{2015}).

\bibitem{2006MNRAS.366..787K}
\bibinfo{author}{{Krajnovi{\'c}}, D.}, \bibinfo{author}{{Cappellari}, M.}, \bibinfo{author}{{de Zeeuw}, P.~T.} \& \bibinfo{author}{{Copin}, Y.}
\newblock \bibinfo{title}{{Kinemetry: a generalization of photometry to the higher moments of the line-of-sight velocity distribution}}.
\newblock \textit{\bibinfo{journal}{Mon. Not. R. Astron. Soc.}} \textbf{\bibinfo{volume}{366}}, \bibinfo{pages}{787--802} (\bibinfo{year}{2006}).

\bibitem{2005MNRAS.356.1293Y}
\bibinfo{author}{{Yang}, X.}, \bibinfo{author}{{Mo}, H.~J.}, \bibinfo{author}{{van den Bosch}, F.~C.} \& \bibinfo{author}{{Jing}, Y.~P.}
\newblock \bibinfo{title}{{A halo-based galaxy group finder: calibration and application to the 2dFGRS}}.
\newblock \textit{\bibinfo{journal}{Mon. Not. R. Astron. Soc.}} \textbf{\bibinfo{volume}{356}}, \bibinfo{pages}{1293--1307} (\bibinfo{year}{2005}).

\bibitem{2007ApJ...671..153Y}
\bibinfo{author}{{Yang}, X. et~al.}
\newblock \bibinfo{title}{{Galaxy Groups in the SDSS DR4. I. The Catalog and Basic Properties}}.
\newblock \textit{\bibinfo{journal}{Astrophys. J.}} \textbf{\bibinfo{volume}{671}}, \bibinfo{pages}{153--170} (\bibinfo{year}{2007}).

\bibitem{2024PhRvD.110b3512L}
\bibinfo{author}{{Li}, S.}, \bibinfo{author}{{Sheng}, M.-J.}, \bibinfo{author}{{Li}, H.} \& \bibinfo{author}{{Yu}, H.-R.}
\newblock \bibinfo{title}{{Lagrangian space remapping and the angular momentum reconstruction from cosmic structures}}.
\newblock \textit{\bibinfo{journal}{Phys. Rev. D}} \textbf{\bibinfo{volume}{110}}, \bibinfo{pages}{023512} (\bibinfo{year}{2024}).

\bibitem{2022MNRAS.509.4024D}
\bibinfo{author}{{Dom{\'\i}nguez S{\'a}nchez}, H.}, \bibinfo{author}{{Margalef}, B.}, \bibinfo{author}{{Bernardi}, M.} \& \bibinfo{author}{{Huertas-Company}, M.}
\newblock \bibinfo{title}{{SDSS-IV DR17: final release of MaNGA PyMorph photometric and deep-learning morphological catalogues}}.
\newblock \textit{\bibinfo{journal}{Mon. Not. R. Astron. Soc.}} \textbf{\bibinfo{volume}{509}}, \bibinfo{pages}{4024--4036} (\bibinfo{year}{2022}).

\bibitem{2025ApJ...992L..17W}
\bibinfo{author}{{Wang}, P.}
\newblock \bibinfo{title}{{Observational Evidence for Spin Alignment between Galaxy Groups and Their Central Galaxies}}.
\newblock \textit{\bibinfo{journal}{Astrophys. J. Lett.}} \textbf{\bibinfo{volume}{992}}, \bibinfo{pages}{L17} (\bibinfo{year}{2025}).

\bibitem{2015ApJ...798...17Z}
\bibinfo{author}{{Zhang}, Y. et~al.}
\newblock \bibinfo{title}{{Spin Alignments of Spiral Galaxies within the Large-scale Structure from SDSS DR7}}.
\newblock \textit{\bibinfo{journal}{Astrophys. J.}} \textbf{\bibinfo{volume}{798}}, \bibinfo{pages}{17} (\bibinfo{year}{2015}).

\bibitem{2013arXiv1308.0847L}
\bibinfo{author}{{Levi}, M. et~al.}
\newblock \bibinfo{title}{{The DESI Experiment, a whitepaper for Snowmass 2013}}.
\newblock \textit{\bibinfo{journal}{arXiv e-prints}} \bibinfo{pages}{arXiv:1308.0847} (\bibinfo{year}{2013}).

\bibitem{2017AJ....154...28B}
\bibinfo{author}{{Blanton}, M.~R. et~al.}
\newblock \bibinfo{title}{{Sloan Digital Sky Survey IV: Mapping the Milky Way, Nearby Galaxies, and the Distant Universe}}.
\newblock \textit{\bibinfo{journal}{Astron. J.}} \textbf{\bibinfo{volume}{154}}, \bibinfo{pages}{28} (\bibinfo{year}{2017}).

\bibitem{2022MNRAS.515.5081Z}
\bibinfo{author}{{Zhou}, Y. et~al.}
\newblock \bibinfo{title}{{SDSS-IV MaNGA: global properties of kinematically misaligned galaxies}}.
\newblock \textit{\bibinfo{journal}{Mon. Not. R. Astron. Soc.}} \textbf{\bibinfo{volume}{515}}, \bibinfo{pages}{5081--5093} (\bibinfo{year}{2022}).

\bibitem{2025ApJ...982L..29B}
\bibinfo{author}{{Bao}, M. et~al.}
\newblock \bibinfo{title}{{The Large-scale Structure Supplies the Formation of Gas-star Misaligned Galaxies}}.
\newblock \textit{\bibinfo{journal}{Astrophys. J. Lett.}} \textbf{\bibinfo{volume}{982}}, \bibinfo{pages}{L29} (\bibinfo{year}{2025}).

\bibitem{2018ApJS..237...24Y}
\bibinfo{author}{{Yu}, H.-R.}, \bibinfo{author}{{Pen}, U.-L.} \& \bibinfo{author}{{Wang}, X.}
\newblock \bibinfo{title}{{CUBE: An Information-optimized Parallel Cosmological N-body Algorithm}}.
\newblock \textit{\bibinfo{journal}{Astrophys. J. Suppl.}} \textbf{\bibinfo{volume}{237}}, \bibinfo{pages}{24} (\bibinfo{year}{2018}).

\bibitem{yu_cube2_2026}
\bibinfo{author}{Yu, H.-R. et~al.}
\newblock \bibinfo{title}{{CUBE2}: {A} parallel {N}-body simulation code for scalability, accuracy, and memory efficiency}.
\newblock \textit{\bibinfo{journal}{Sci. China Phys. Mech. Astron.}} \textbf{\bibinfo{volume}{69}}, \bibinfo{pages}{269511} (\bibinfo{year}{2026}).

\bibitem{2009ApJS..180..306D}
\bibinfo{author}{{Dunkley}, J. et~al.}
\newblock \bibinfo{title}{{Five-Year Wilkinson Microwave Anisotropy Probe Observations: Likelihoods and Parameters from the WMAP Data}}.
\newblock \textit{\bibinfo{journal}{Astrophys. J. Suppl.}} \textbf{\bibinfo{volume}{180}}, \bibinfo{pages}{306--329} (\bibinfo{year}{2009}).

\bibitem{1970A&A.....5...84Z}
\bibinfo{author}{{Zel'dovich}, Y.~B.}
\newblock \bibinfo{title}{{Gravitational instability: An approximate theory for large density perturbations.}}
\newblock \textit{\bibinfo{journal}{Astron. Astrophys.}} \textbf{\bibinfo{volume}{5}}, \bibinfo{pages}{84--89} (\bibinfo{year}{1970}).

\bibitem{1981csup.book.....H}
\bibinfo{author}{Hockney, R.~W.} \& \bibinfo{author}{Eastwood, J.~W.}
\newblock \bibinfo{title}{{Computer Simulation Using Particles.}}
\newblock \bibinfo{publisher}{McGraw-Hill}, \bibinfo{address}{New York} (\bibinfo{year}{1981}).

\bibitem{sheng_2026_20263913}
\bibinfo{author}{Sheng, M.-J. et~al.}
\newblock \bibinfo{title}{{Observed galaxy spin and spin reconstructed from cosmic initial condition.}}
\newblock \textit{\bibinfo{journal}{Zenodo}} \textbf{\bibinfo{volume}{20263913}} (\bibinfo{year}{2026}).

\bibitem{2026ascl.soft04001Y}
\bibinfo{author}{{Yu}, H.-R. et~al.}
\newblock \bibinfo{title}{{CUBE2: Optimized cosmological N-body simulation code.}}
\newblock \textit{\bibinfo{journal}{Astrophysics Source Code Library}} \textbf{\bibinfo{volume}{ascl:2604.001}} (\bibinfo{year}{2026}).

\end{thebibliography}
\end{document}